\documentclass[preprint,review,12pt]{elsarticle}
\usepackage{lineno,hyperref}
\usepackage{color}
\usepackage{amsmath}
\usepackage{bm}
\usepackage{setspace}   

\usepackage{float}

\journal{International Journal of Mechanical Sciences}

\graphicspath{{Manuscript/}}

\usepackage{float}
\usepackage{xr}
\usepackage{sidecap}
\usepackage{graphicx,txfonts}

\usepackage{tabularx}

\usepackage{algorithm}
\usepackage{algorithmic}

\biboptions{numbers,sort&compress}

\begin{document}

\begin{frontmatter}

\title{Exploring the design space for nonlinear buckling of \\
\bigskip
composite thin-walled lenticular tubes under pure bending}




\bigskip

\author[aff1]{Qilong Jia}
\author[aff1]{Ning An*}
\ead{anning003@stu.xjtu.edu.cn}
\author[aff2]{Xiaofei Ma}
\author[aff1]{Jinxiong Zhou}
\cortext[corr]{Corresponding author}

\bigskip

\address[aff1]{State Key Laboratory for Strength and Vibration of Mechanical Structures and School of \\
\medskip
Aerospace, Xi'an Jiaotong University, Xi'an 710049, People’s Republic of China\\
\medskip}

\address[aff2]{Xi’an Institute of Space Radio Technology, Xi’an 710100, People’s Republic of China}

\begin{abstract}

This paper presents an automatic finite element simulation scheme accounting for high geometric nonlinearity and the difference between linear and nonlinear buckling of composite thin-walled lenticular tubes (CTLTs). Parameterizing of cross-section shapes and generation of design space for CTLTs with both circular and parabolic arcs were accomplished, and several key factors were identified, in particular the contrary effect of lumbus length and parabolic coefficient on the bending stiffness anisotropy. The first quantitative comparison of triangular rollable and collapsible (TRAC) booms and CTLTs is given in terms of bending performance in two directions, showing that the optimal CTLT carefully selected from the design space demonstrates a comparable or even better performance than the TRAC boom. This is of great importance from both academic and engineering perspectives. Our efforts enhance the understanding of nonlinear buckling and post-buckling behavior of CTLTs, and provide guidelines for future design of CTLTs with desirable performance.
\end{abstract}

\begin{keyword}
Nonlinear buckling; Composite thin-walled lenticular tubes (CTLTs); Finite element method; Pure bending.

\end{keyword}

\end{frontmatter}



\section{Introduction}


The composite thin-walled lenticular tube (CTLT), which is also known as the collapsible tube mast (CTM), is a closed tube with lenticular cross-section that can be commonly used for deploying large structures in space. The CTLT is flattened and coiled for storage before and during launch, and capable of deploying spontaneously for use once in orbit. This type of structure, since was first introduced by NASA a few decades ago~\cite{rennie1967new}, has been developed and employed as a basic component to deploy various large space structures, such as solar arrays and antennas in many space missions. The first CTM ever used in space was jointly developed by NASA and ESA for the ULYSSES mission~\cite{aguirre1985collapsible}. The development of CTLT is also considered as a key technique of German Aerospace Center DLR's solar sailing technology~\cite{leipold2005large,block2011ultralight,hillebrandt2014boom,belvin2016advanced}, and in 2009 an agreement was made between ESA and DLR by which they started a three-step project aiming to develop, prove, and demonstrate that CTLT can serve as a safe and reliable component for long-lasting and deep space missions~\cite{geppert20113,seefeldt2017gossamer}. NASA has also recently expressed an interest in small CTLTs as a candidate solar sail boom for low-cost deep space exploration and science missions~\cite{fernandez2017advanced,fernandez2018advanced,firth2019advanced,firth2020minimal,long2021multiscale,stohlman2021characterization}. As an example of commercial application, Oxford Space Systems Ltd (OSS) in UK, is currently developing a 2.7-m-diameter wrapped-rib antenna where 48 CTLTs are employed as wrapping ribs to deploy the metal mesh reflector surface~\cite{angevain2019large,yoshiro2019deployable,curiel2019synthetic}. 

The CTLT may be subjected to various kinds of mechanical loads during different stages of its working process such as flattening, coiling and deploying, and as a slender thin-walled structure it could demonstrate a complex and nonlinear behavior. A large number of fundamental research has been carried out to study the flattening and wrapping process of CTLTs. Hu et al.~\cite{hu2017study} performed a combined experimental, numerical, and analytical investigation of the mechanical response of both compressive and tensile flattening deformations of deployable CTLTs. Chen et al.~\cite{chen2017experimental} carried out some experiments to test the large deformation behaviors of CTLTs in flattening and wrapping process and developed three-dimensional finite element models to predict the mechanical characteristics identified by experiments. Similar studies were also reported by Bai et al.~\cite{bai2013analytical,bai2017thermal,bai2019folding} for determining tensile, compression and folding behaviors of CTLTs. The above-mentioned efforts have been focused on the deformation and the associated failure during flattening and folding, and localized buckling that occurs during flattening and folding was observed as the dominant failure mode of deployable CTLTs~\cite{sickinger2006structural,bai2014temperature}. However, little attention was paid to study the instability or buckling behavior of CTLTs. The only relevant work that we are aware of focused on the linear buckling and nonlinear post-buckling response of the CTLT subjected to uniaxial compression~\cite{hu2016mechanical}. It was found that the critical buckling load estimated directly by eigenvalue analysis is far greater than the realistic experimental measurement. A closer critical buckling load was eventually obtained by introducing proper initial imperfections and performing a post-buckling analysis. However, the initial imperfections introduced in the post-buckling analysis were determined from the experimental observations and this may reduce the predictive significance of the model~\cite{shirkavand2019orientation}. Therefore, reliable simulation techniques need to be developed to accurately predict the critical buckling load of CTLTs prior to conducting experiments~\cite{devarajan2020thermal}.

A more accurate prediction of the critical buckling load can be obtained by first performing a geometrically nonlinear response analysis and then estimating the buckling load by a following eigenvalue analysis on the deformed configuration. This method is referred to as the nonlinear buckling analysis, and it has been well-developed and documented, and successfully applied in the field of mechanical engineering in the past few decades. A number of analytical solutions can be found for the nonlinear buckling predictions of simple structures such as beam, truss, and shell problems with ideal boundary conditions. Brendel et al.~\cite{brendel1980linear} formulated the eigenvalue problem with the information at deformed states on the nonlinear pre-buckling path and studied the effect of imperfections on the critial loads of several cylindrical shells under uniform pressure and wind load. Kounadis et al.~\cite{kounadis1985efficient} proposed a simplified approach for the nonlinear buckling analysis of frames subjected to either bifurcational or limit-point instability. Wu et al.~\cite{wu1987design,wu1988design} derived the design derivatives of the nonlinear critical load and proposed an optimization procedure to maximize the load-carrying capacity of truss-structures. For complex systems, analytical solutions are rarely obtainable and numerical modelling is often required. Lindgaard et al.~\cite{lindgaard2010nonlinear_1,lindgaard2011optimization,lindgaard2011unified,lindgaard2010nonlinear_2} developed a combination of numerical approach and optimization procedure to maximize the critical nonlinear buckling load of laminated composite shell structures and provided validation with several benchmark problems. Nguyen et al.~\cite{nguyen2013numerical} formulated a finite element model using three-noded Timoshenko beam elements to analyze the nonlinear buckling load of an inflatable beam made of orthotropic technical textiles. Liang et al.~\cite{liang2018nonlinear,liang2021novel} proposed a reduced order model that capable of predicting the nonlinear buckling behavior of variable stiffness composite plates efficiently compared with typical full order finite element model.

Structural instability arises as one of the biggest concerns for structural design of thin-wall composite boom structures such as CTLTs in aerospace engineering, because these structures are usually thin in thickness and have long aspect ratio. TRAC (triangular rollable and collapsible) booms and CTLTs have been considered promising (competitive) candidates for solutions of various light-weight deployable structures~\cite{spencer2019solar,firth2019deployment,sullivan2020boom}. Very recently a few studies have been carried out to understand the nonlinear buckling behavior of TRAC booms subjected to pure bending. Murphey et al.~\cite{murphey2017trac} analyzed the basic structural mechanics including deployment stiffness, buckling strength, and packaging constraints of TRAC booms using closed form analytical and finite element approaches. Leclerc et al.~\cite{leclerc2017characterization} performed a study of the nonlinear elastic buckling behavior of triangular rollable and collapsible (TRAC) booms under pure bending and reported a good agreement between numerical predictions and experimental measurements. Bessa et al.~\cite{bessa2018design} constructed the design space diagram for the TRAC booms and proposed a data-driven computational framework combining Bayesian regression for optimizing the nonlinear critical buckling load of TRAC booms. Despite the available study on instability of TRAC booms and CTLTs, either numerical or experimental, reliable prediction of critical loads with emphasis on variation of configuration and initial stress field in the context of nonlinear buckling analysis remains elusive. A recent study on TRAC booms reveals the subtle difference between classical linear buckling analysis and nonlinear buckling, and the linear buckling may give biased estimation on critical loads~\cite{leclerc2020nonlinear,cox2019scalability}. But this has never been carried out for CTLTs. Moreover, due to the unique cross-section shape of CTLT, it exhibits strong bending stiffness anisotropy~\cite{lee2018mechanics,lee2019inducing,salazar2021experimental,wilkieoverview}, and thus a rational design of CTLT is only achievable provided an accurate design space diagram is available, and a trade-off is made by compromising bending performance in different directions. Another benefit of generating and exploiting through a design space is the possibility of unveiling some key design parameters which might be missed otherwise. Last but not least, the obtained design space and exploration permits parameter tailoring for optimal designs, and more importantly, allows a possible quantitative comparison of the two counterparts, TRAC booms and CTLTs. 

In this paper, we develop a finite element method based numerical scheme that predicts the nonlinear buckling and post-buckling behavior of CTLTs under pure bending. We first study the mechanical response of a particular CTLT bent along two perpendicular directions, and demonstrate the validity of the developed analysis methods. A good agreement was obtained between the predictions of critical buckling load extracted by the nonlinear buckling analysis and that identified from the post-buckling response. We then perform a systematic numerical study to construct the design space against nonlinear buckling for a variety of CTLTs with equal weight but different cross-section shapes. The cross-section shape is shown to be a convenient parameter for controlling the pre-buckling stiffness and the nonlinear critical buckling load of CTLTs. More specifically, we start by investigating the effect of the size of web and lumbus on the bending resistance performance of conventional CTLTs with circular arcs. We then introduce the concept of parabolic CTLTs by replacing the conventional circular arcs with parabolas, and show how it is possible, by just tuning the parabolic shape, to alter the range of attainable mechanical responses and leverage the trade-off between the bending resistance performance in the two directions. Finally, we made a comparison of the bending performance between conventional circular CTLTs, parabolic CTLTs and TRAC booms with equal weight. It is shown that the optimal CTLT carefully selected from the design space demonstrates a comparable or even better performance than TRAC boom loaded concurrently in two loading directions.

The paper is organized as follows. Section~\ref{Prodescrip} describes the geometry design of the traditional CTLTs with circular arcs. Section~\ref{FEM_modeling} introduces the numerical analysis techniques that are used to investigate the nonlinear buckling and post-buckling response of CTLTs. Section~\ref{results} presents the results obtained from parametric studies, highlighting the effect of cross-section shape on the nonlinear mechanical response of CTLTs. Finally, concluding remarks are included in Section~\ref{conclusion}.

\section{CTLT Geometry}
\label{Prodescrip}

\begin{figure}[h]
\centering
\includegraphics[width=140mm]{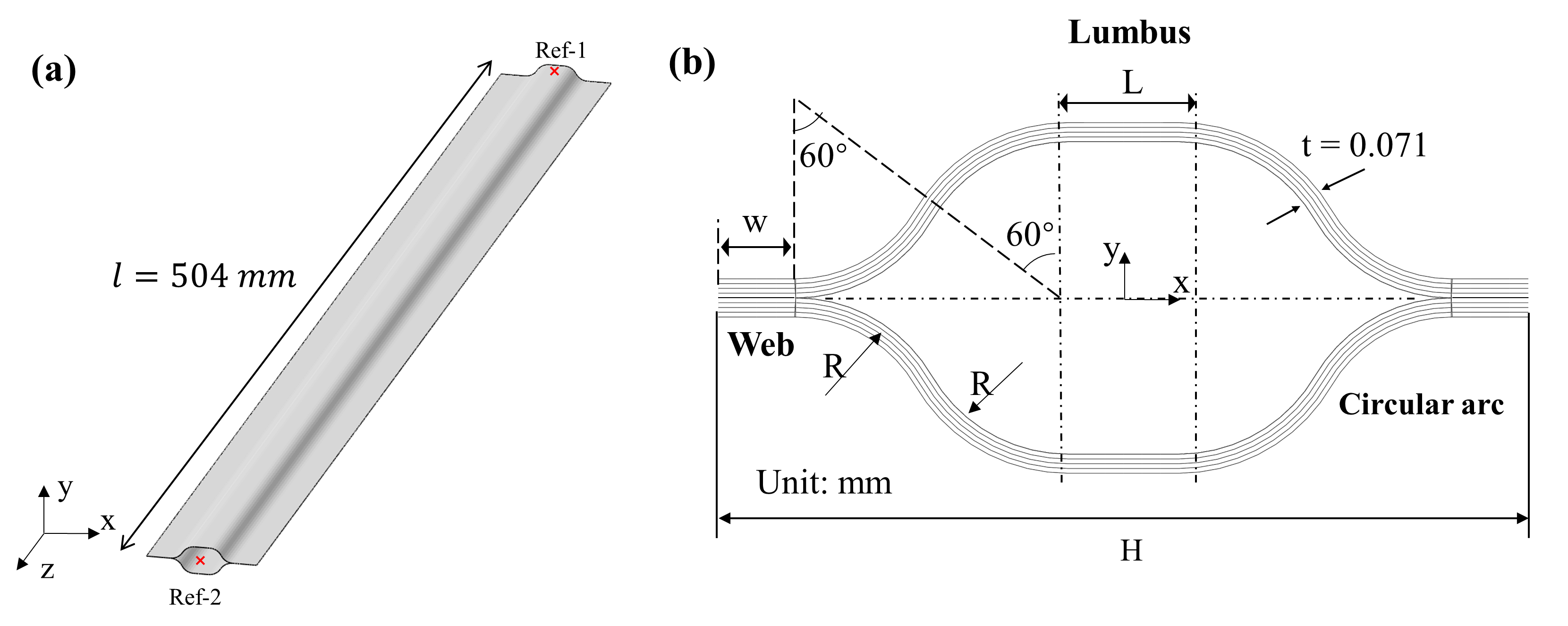}
\caption{Schematic of the CTLT with circular arcs. (a) A CTLT in the deployed, i.e., unstressed configuration. (b) Cross-section of the CTLT with geometric parameters indicated. Web width $w$ and lumbus length $L$ are considered as the two independent variables, and circular arc radius $R$ is found from the flattening length equation.}
\label{Figure1}
\end{figure}

Fig. \ref{Figure1} shows a schematic of the conventional CTLT with circular arcs. This type of structure consists of joining two thin omega-shaped cylindrical shells of thickness $t$. The cross-section of each omega-shaped shell consists of four circular arcs of radius $R$ subtending an angle $60^{\circ}$, two straight segments at the ends of width $w$ constituting the web, and one straight segment called lumbus of length $L$ at center~\cite{royer2018ultralight}. The two shells are mirror-symmetric and the end flat parts are bonded together in the unstressed configuration. The geometry of the CTLT is then fully characterized by the longitudinal length $l$, thin-shell thickness $t$, and the cross-section parameters: web width $w$, lumbus length $L$, and circular arc radius $R$. In order to make a fair comparison between the performance of CTLTs with different cross-section geometries and TRAC booms, the longitudinal length $l$, thin-shell thickness $t$, and the flattening length of the cross-section $s$ are set the same as those of TRAC booms reported in literature~\cite{leclerc2017characterization, bessa2018design}, i.e., $l = 504$ mm, $t=0.071$ mm and $s = 27.43$ mm. Finally, the flattening length of the cross-section $s$, which can be expressed as $s = 2w + L + \frac{4}{3}{\pi}R$, is fixed constant, and therefore the mass of the structure remains unchanged. The web width $w$ and the lumbus length $L$ are selected as two independent variables that are used to adjust the cross-section geometry of circular CTLTs.

\section{Finite element analyses}
\label{FEM_modeling}

In this section, we will detail the finite element analyses that are used to investigate the nonlinear buckling and post-buckling behavior of CTLTs subjected to pure bending. First, we will introduce the finite element model with the definition of boundary conditions and material properties (see Section~\ref{Finite element model}). Then, we will focus on the analyses to predict the critical loads for the onset of buckling and the associated buckling modes  (see Section~\ref{Instability analysis}). Finally, we will discuss the post-buckling response analysis (see Section~\ref{Post-buckling analysis}).

\subsection{Finite element model}
\label{Finite element model}

The finite element model is constructed utilizing the commercial software Abaqus 2020 with python scripts allows for automated parametric studies. Four-node general-purpose shell elements with reduced integration (Abaqus element type S4R) were used and the accuracy of the mesh was ascertained through a mesh refinement study, resulting in a relative mesh density of around 50 elements along the cross-section profile.

The nodes forming the two end cross-sections are kinematically coupled to two reference points, in effect creating rigid cross-sections that match the end conditions of the experimental setup~\cite{leclerc2017characterization}. All six degrees of freedom of the reference point at end 1 are constrained, defining a clamped condition. At the other end, a pure moment load is applied to the reference point 2. The location of the two reference points is shown in Fig.~\ref{Figure1}(a). In the case of a moment applied about $x-$axis $M_{X}$, the translational degrees of freedom along $Y$ and $Z$ as well as the rotational degree of freedom around $X$ are left free, while the other three degrees of freedom are all fixed, i.e., $u1=$ 0, $u2=$ FREE, $u3=$ FREE, $ur1=$ FREE, $ur2=$ 0, $ur3=$ 0. In the case of a moment applied about $y-$axis $M_{Y}$, the same boundary conditions are used with $X$ and $Y$ inverted.

In accordance with previously reported data for TRAC booms~\cite{leclerc2017characterization,bessa2018design}, each CTLT omega shell consists of four layers of unidirectional carbon fibers in an epoxy resin and the layers are arranged in the stacking sequence $[0^{\circ}, 90^{\circ}]_S$. The nominal orthotropic elastic material properties of each layer are set as $E_{1}=128.0$ GPa, $E_{2}=6.5$ GPa, $\nu_{12}=0.35$, $G_{12}=G_{13}=G_{23}=7.5$ GPa. In addition, a composite shell section is defined to specify the material properties, thickness, and orientation angle of each layer.

\subsection{Instability analysis}
\label{Instability analysis}

In a nonlinear finite element formulation, the response of a structure is calculated by iteratively solving the equation~\cite{bathe1983automatic,wriggers2008nonlinear,kim2014introduction}:
\begin{equation}
    K_T \delta u = R
    \label{eq1}
\end{equation}
where $K_T$ is the current tangent stiffness matrix where the loads are applied, $\delta u$ is the incremental displacement vector, and $R$ is the force residual for the iteration. For large deflection problems, the tangent stiffness matrix $K_T$ consists of the sum of the initial stiffness matrix $K_0$, the displacement stiffness matrix $K_L$, and the stress stiffness matrix $K_\sigma$ as described by:
\begin{equation}
    K_T  = K_0 + K_L + K_\sigma
    \label{eq2}
\end{equation}
The structure is stable only if the tangent stiffness matrix $K_T$ is positive definite; in other words, the structure becomes unstable and buckling is likely to occur when the tangent stiffness $K_T$ is singular. Therefore, the onset of buckling can be predicted by looking for the loads for which the model tangent stiffness matrix becomes singular, and so that the problem
\begin{equation}
K_T \delta u=0
\label{eq3}
\end{equation}
has non-trivial incremental displacement vector $\delta u$ solutions.

\subsubsection{Linear buckling analysis}

In the linear buckling analysis, the pre-buckling displacement is assumed to be small thus the displacement stiffness matrix $K_L=0$, and the tangent stiffness matrix is approximated using only the initial stiffness matrix $K_0$ and the stress stiffness matrix $K_\sigma$. By assuming the stress stiffness matrix to be linearly proportional to an incremental load $Q$, Eq.~\ref{eq3} is reduced to an eigenvalue problem:
\begin{equation}
    (K_{0} + \lambda_{i}K_{\sigma}) \phi_{i}=0
\end{equation}
where $K_0$ is the initial global stiffness matrix defined at the undeformed configuration, $K_\sigma$ is the initial stress stiffness matrix caused by the incremental load $Q$, $\lambda_{i}$ are the eigenvalues, i.e., critical buckling load factors, and  $\phi_{i}$ are the eigenvectors, i.e., buckling modes. The eigenvalues and eigenvectors are ordered increasingly in magnitude, such that $\lambda_{1}Q$ is the lowest linear critical buckling load and $\phi_{1}$ is the corresponding linear buckling mode. This eigenvalue problem is readily solved using Lanczos method by an eigenvalue buckling analysis step (*Buckle) in Abaqus.

\subsubsection{Nonlinear buckling analysis}
In the nonlinear buckling analysis, the effect of the change in configuration in the pre-buckling stage is considered and the full tangent stiffness matrix from Eq.~\ref{eq2} must be utilized~\cite{pedersen2018buckling}. The nonlinear buckling analysis generally consists of two analysis steps: (\emph{i}) a nonlinear static analysis step (*Static) and (\emph{ii}) an eigenvalue buckling analysis step (*Buckle). In the first step, a preload ("dead" load), $P$, is applied to the structure to attain a deformed state, during which the large-displacement formulation is used (NLGEOM=ON) to capture the geometric nonlinearity. Then the deformed state of the model at the end of the first step is identified as the base state for the second eigenvalue buckling step. Namely, the nonlinear critical loads and buckling modes are calculated by applying a seperate incremental load $Q$ to the deformed state of the structure with change of configurations and attained stress-field incorporated. The eigenvalue problem is a bit complicated as follows:
\begin{equation}
    (K_{0} + K_{L} + \lambda_{i}K_{\sigma}) \phi_{i}=0
\end{equation}
where $K_{0}$ is the initial global stiffness matrix defined at the undeformed configuration, $K_{L}$ is the global displacement stiffness matrix determined at the deformed state by applying the preload $P$, and $K_\sigma$ is the initial stress stiffness matrix caused by the incremental load $Q$.

While large deformation is included in the static analysis, the eigenvalue buckling theory relies on there being little geometric change due to the "live" buckling load, $\lambda_{i}Q$. Then the final nonlinear critical loads predicted by the two-step nonlinear buckling analysis are given by $P+\lambda_{i}Q$ and $\phi_i$ denote the corresponding nonlinear buckling modes. Notice that the amount of preload would also have an effect on the nonlinear buckling predictions of structures. To capture the nonlinear buckling behavior of the thin-walled composite structures which are known for their high sensitivity to geometric imperfections and having many buckling modes with closely spaced eigenvalues, it often helps to apply enough preload to deform the structure to just below the buckling load prior to performing the eigenvalue extraction~\cite{zheng2009wrinkling}. On the other hand, the structure should not be preloaded above the buckling load; otherwise the adopted Lanczos algorithm in Abaqus will issue an error message and terminate the analysis. 

In the nonlinear buckling analysis, with the aim of finding the appropriate preload which should be as large as possible while allowing the extraction of eigenvalues, the following tricks could be helpful: (\emph{i}) applying a relatively large preload (greater than nonlinear buckling load) in the nonlinear static analysis step, and thus this step is expected to be conducted up to the predetermined load or fail to converge due to buckling; and (\emph{ii}) performing an eigenvalue analysis in the deformed state starting from the gradually decreasing last available increment until the eigenvalues can be properly extracted. The above procedure, illustrated as pseudocode in Algorithm~\ref{algorithm1}, is implemented into a Python script in Abaqus to run the simulations automatically. In real practice, the preload in the nonlinear static analysis is set equal to the linear critical buckling load obtained by conducting a linear buckling analysis, because the linear buckling analysis never underestimates the critical load of the CTLT as will be demonstrated in the following sections.

\begin{algorithm}
\caption{Pseudocode for the nonlinear buckling analysis}
\label{algorithm1}
\begin{algorithmic}
\STATE Step 1. Run a nonlinear static analysis with a preload being applied and write the model definition and deformed state at every increment to the files required for restart.
\STATE Step 2. Perform an eigenvalue analysis from the last available increment until the eigenvalues can be successfully extracted.
\WHILE{True:}
\STATE Perform the eigenvalue analysis
\IF{Eigenvalue analysis completes successfully}
\STATE break
\ENDIF
\STATE n = n - 1 (n is the maximum number of available increments obtained in Step 1)
\ENDWHILE
\end{algorithmic}
\end{algorithm}

\subsection{Post-buckling analysis}
\label{Post-buckling analysis}
The nonlinear post-buckling response of the CTLT is investigated by introducing an imperfection in the form of the two most critical buckling modes, $v^M_1$ and $v^M_2$, obtained from the nonlinear buckling analysis. Therefore, the mesh is perturbed by $v^M_1$ and $v^M_2$ scaled by a factor $\eta$, such that
\begin{equation}
    {\delta}v^M = \eta(v^M_1 + v^M_2)
\end{equation}
where $\eta$ is chosen as 5\% of thickness of the CTLT, i.e., $\eta=0.05t$.

\section{Results and discussions}
\label{results}

In this study numerical simulations are performed to explore the nonlinear buckling behavior space of CTLTs characterized by different cross-section geometries. First, the linear and nonlinear buckling and post-buckling behavior of a particular CTLT is investigated (see Section~\ref{Behavior_A_CTLT}). Then, the behavior space of circular CTLTs is explored, highlighting the effect of the size of web and lumbus on the pre-buckling stiffness and nonlinear critical buckling load of conventional CTLTs (see Section~\ref{BehaviorSpace_CircularCTLTs}). Next, the concept of parabolic CTLTs is introduced and the design space of which is also probed, demonstrating a wider tunability range of the mechanical response (see Section~\ref{BehaviorSpace_ParabolicCTLTs}). Finally, the bending resistance performance of both circular and parabolic CTLTs is compared to that of TRAC booms with equal weight, and a discussion is presented (see Section~\ref{Comparison}).

\subsection{Nonlinear buckling behavior of a particular circular CTLT}
\label{Behavior_A_CTLT}

A beam develops compressive stresses on its inner surface when it is subjected to a pure bending moment. For the CTLTs, the bending moment about $x-$axis produces in-plane compressive stresses on the inner lumbus, while the bending moment about $y-$axis produces in-plane compressive stresses on the inner web. The compressive stress is the driving force for most buckling phenomena. Moreover, the lumbus and web, as being parts of the CTLT, are thin-walled structures and have a low bending stiffness. Therefore, the in-plane compressive stress may lead to out-of-plane buckling within the lumbus or web when the critical value is reached.

\begin{figure}[h]
\centering
\includegraphics[width=150mm]{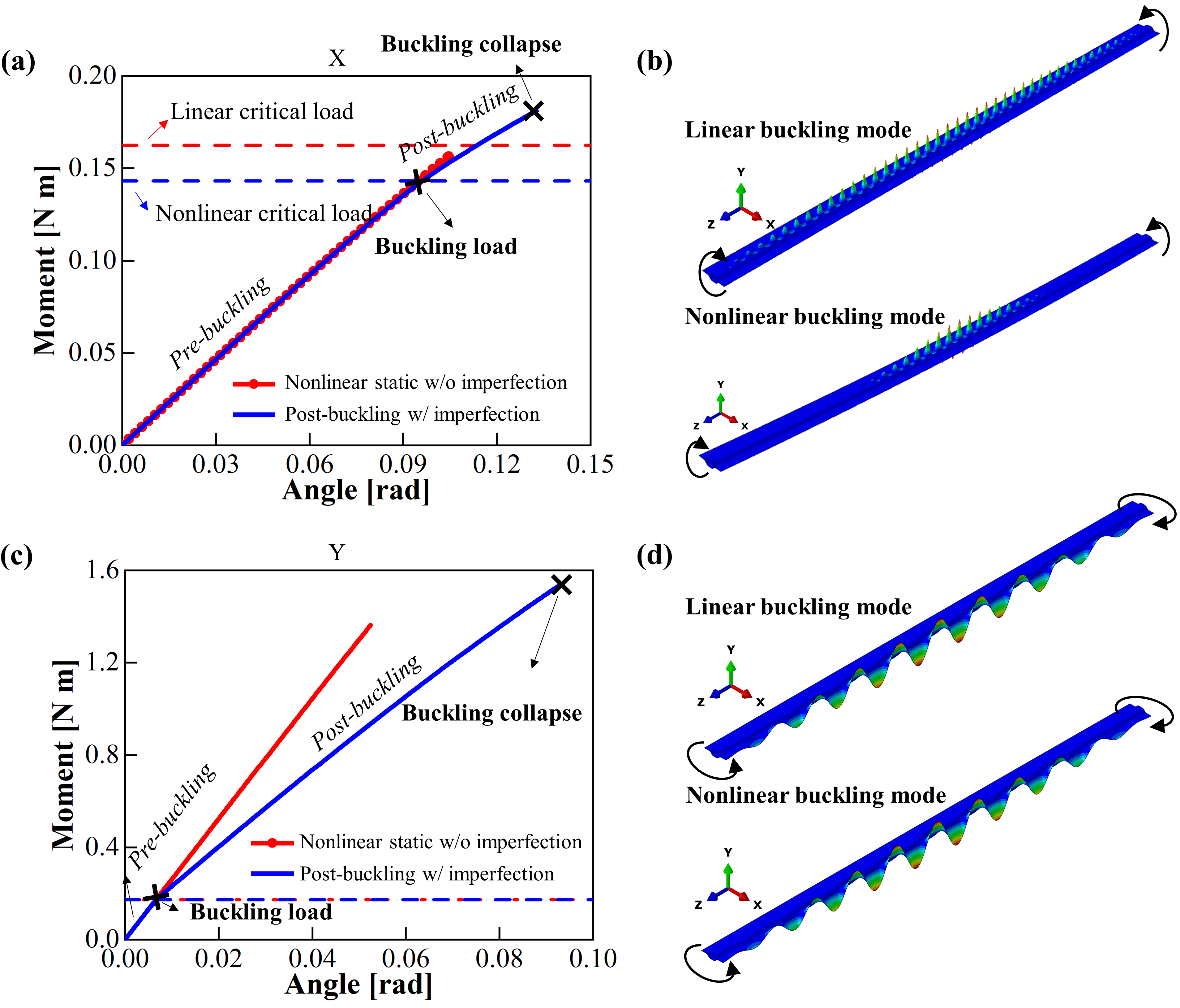}
\caption{Nonlinear buckling and post-buckling response of a particular CTLT characterized by $w=5.5$ mm and $L=3.5$ mm when subjected to a pure bending moment about (a-b) $x-$ axis and (c-d) $y-$axis.}
\label{Figure2}
\end{figure}

To demonstrate the typical response characteristics of the CTLTs under pure bending, without loss of generality, in Fig.~\ref{Figure2} we present the analysis results of a particular CTLT characterized by $w=5.5$ mm and $L=3.5$ mm. Fig.~\ref{Figure2}a and ~\ref{Figure2}b show the mechanical response of the CTLT when the bending moment is applied about $x-$axis. We start by determining the critical loads for buckling of this CTLT, only the lowest critical load is of interest here. The critical load is determined from different perspectives. First, we estimate the critical load by solving the eigenvalue problem through performing a liner buckling analysis or a nonlinear buckling analysis as discussed previously. The results are presented by two dashed lines perpendicular to the moment axis. In Fig.~\ref{Figure2}a the red dashed line indicates the critical load obtained from the linear buckling analysis, which we refer to as \emph{linear critical load (LCL)}, and the blue dashed line indicates that obtained from the nonlinear buckling analysis, which we refer to as \emph{nonlinear critical load (NLCL)}. Moreover, the critical buckling load can also be identified from the moment-angle curve by performing a post-buckling analysis. For the post-buckling analysis, the imperfections afore-mentioned should be introduced. It can be seen from Fig.~\ref{Figure2}a that there exist two distinct regimes in the moment-angle curve (as shown in blue solid line), i.e., a pre-buckling regime and a post-buckling regime. The transition point on the moment-angle curve from pre-buckling regime to post-buckling regime can be identified as the critical load for buckling. From these results shown in Fig.~\ref{Figure2}a, we conclude that (\emph{i}) linear buckling analysis would overestimate the overall stability of the structure in this case, i.e., $LCT > NLCT$, and (\emph{ii}) the nonlinear critical load ($NLCL$) estimated by the nonlinear buckling analysis agrees very well with that identified from the post-buckling response. Furthermore, the difference of linear and nonlinear buckling analysis is also manifested in terms of buckling modes. As shown in Fig.~\ref{Figure2}b, linear buckling analysis predicts a larger number of winkles within the lumbus than nonlinear buckling analysis in this case. It should be pointed out the difference between linear and nonlinear buckling highlighted herein is firstly pointed out by Leclerc et al.~\cite{leclerc2020nonlinear}in the analysis of TRAC boom. The difference is elucidated once again for CTLT bent about $x-$axis. 

In sharp contrast, a different buckling behavior is observed when the pure bending moment is loaded about $y-$axis. It can be seen in Fig.~\ref{Figure2}c that for this case the critical loads predicted by linear and nonlinear buckling analysis are basically the same, i.e., $LCL = NLCL$, and both are in good agreement with that identified from the post-buckling moment-angle curve. This difference is attributed to the different boundary conditions of the web and lumbus in compression. More specifically, when the bending moment is applied about $x-$axis the lumbus undergoing compressive stresses is constrained on both sides by surrounding structures; when the bending moment is applied about $y-$axis, only one side of the web in compression is constrained while the inner edge remains free, which reduces the sensitivity of buckling load to geometric nonlinearity. On the other hand, the post-buckling response under the two loading conditions exhibits similar features, as they both contain two distinct stages divided by $NLCL$, namely, a pre-buckling stage starting from initiation to $NLCL$, and a post-buckling stage starting from $NLCL$ and ending with buckling collapse. We also show, in Fig.~\ref{Figure2}a and~\ref{Figure2}c, that the pre-buckling regime predicted by the post-buckling analysis is observed coincided with the curve obtained by performing a nonlinear static analysis without introducing any imperfections, which exhibits approximately a linear behavior. Then the pre-buckling stiffness of CTLTs can be calculated as
\begin{equation}
    K = \frac{dM}{d\theta}
\end{equation}
where the derivative can be directly obtained through linear regression of the data in pre-buckling regime.

\subsection{Construction of design space for circular CTLTs}
\label{BehaviorSpace_CircularCTLTs}
\begin{figure}[H]
\centering
\includegraphics[width=150mm]{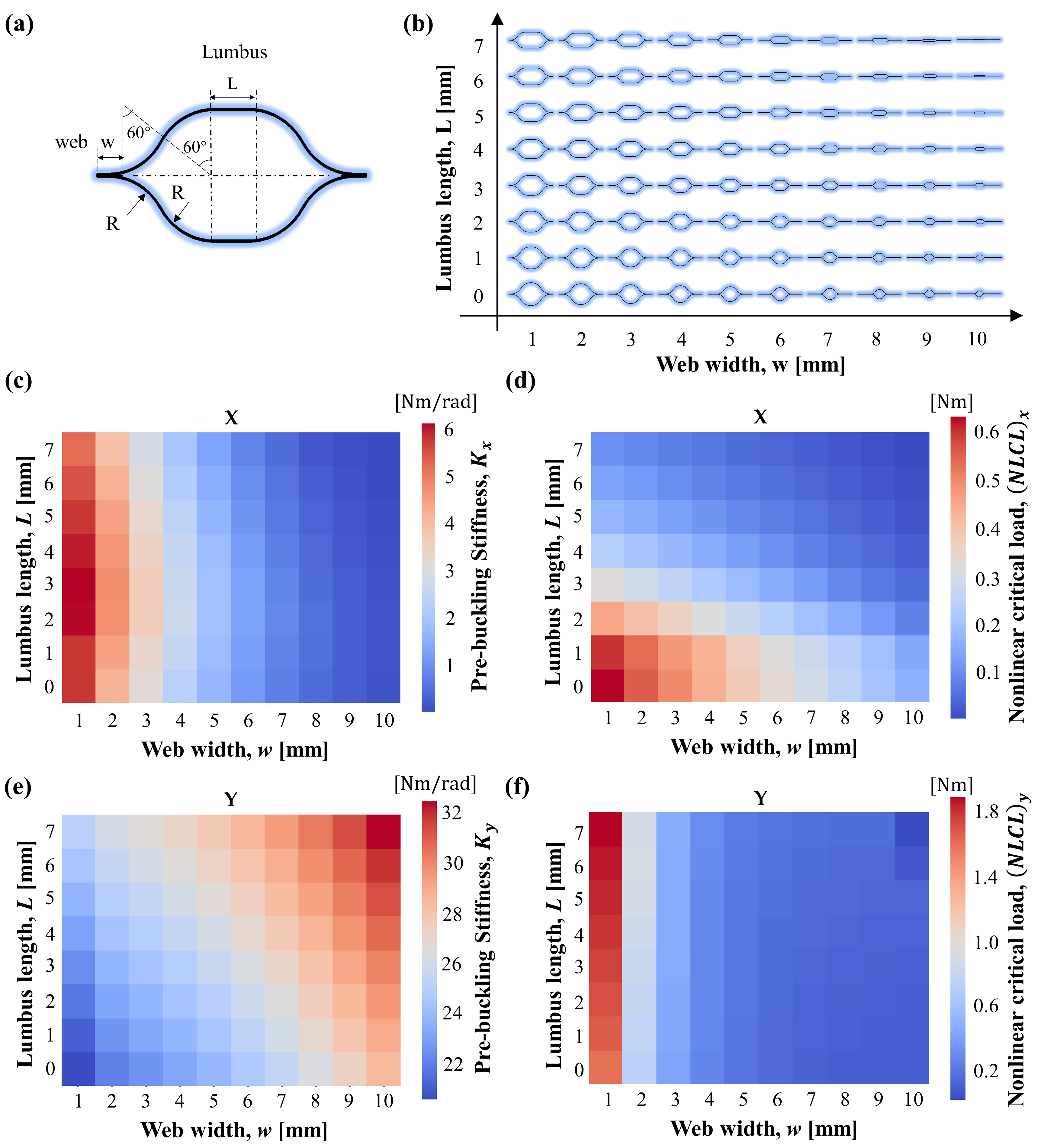}
\caption{The design space for nonlinear buckling of CTLTs with circular arcs. (a) Schematic illustration of the cross-section geometry of the circular CTLT. (b) A variety of cross-section shapes of circular CTLTs characterized by web width $w\in[1, 10]$ mm and lumbus length $L\in[0, 7]$ mm. Heat map illustrating the (c) pre-buckling stiffness $K_x$ and (d) nonlinear critical load $(NLCL)_x$ as a function of the web width $w$ and lumbus length $L$ when the moment is applied about $x-$axis. Heat map illustrating the (e) pre-buckling stiffness $K_y$ (f) and nonlinear critical load $(NLCL)_y$ as a function of the web width $w$ and lumbus length $L$ when the moment is applied about $y-$axis.}
\label{Figure3}
\end{figure}

Having identified typical mechanical behavior of CTLTs under pure bending, we then move on to construct the design space, aiming to probe the evolution of pre-buckling stiffness $K$ and nonlinear critical buckling load $NLCL$. The design space is constructed by varying the two independent cross-section geometric parameters, which are the web width $w$ and lumbus length $L$ as shown in Fig.~\ref{Figure3}a. More specifically, $w$ is varied from 1 to 10 mm and $L$ is varied from 0 to 7 mm, and a number of diverse cross-section shapes are determined as shown in Fig.~\ref{Figure3}b. Note that $L=0$ indicates a CTLT design without lumbus. Fig.~\ref{Figure3}c and ~\ref{Figure3}d present the evolution of the pre-buckling stiffness $K_x$ and the nonlinear critical load $(NLCL)_x$ about $x-$axis as a function of the web width $w$ and the lumbus length $L$. It is shown that the pre-buckling stiffness $K_{x}$ strongly depends on the web width $w$ but slightly on the lumbus length $L$, while the nonlinear critical load $(NLCL)_{x}$ depends both highly on the web width $w$ and the lumbus length $L$. Specifically, $K_{x}$ increases as $w$ decreases but tends to remain unchanged as $L$ varies, while $(NLCL)_x$ increases as $w$ and/or $L$ decreases. Furthermore, the evolution of the pre-buckling stiffness $K_y$ and nonlinear critical load $(NLCL)_y$ about $y-$axis is presented in Fig.~\ref{Figure3}e and~\ref{Figure3}f. It is shown that the pre-buckling stiffness $K_y$ depends both highly on the web width $w$ and the lumbus length $L$, while the nonlinear critical load $(NLCL)_y$ depends solely on the web width $w$. Specifically, $K_y$ increases with the increase of $w$ or $L$, while $(NLCL)_y$ increases with the decrease of $w$. In conclusion, there exists a common region of the design space that maximize the pre-buckling stiffness about $x-$axis and nonlinear critical buckling load about both $x$ and $y-$axes at small web width, although a decrease in the pre-buckling stiffness about $y-$axis is expected. Given the fact that the minimum pre-buckling stiffness about $y-$axis ($\sim$20 Nm/rad) is yet much greater than the maximum pre-buckling stiffness about $x-$axis ($\sim$6 Nm/rad), the smallest value of 1 mm is determined to be the optimal size for the web.

\begin{figure}[H]
\centering
\includegraphics[width=150mm]{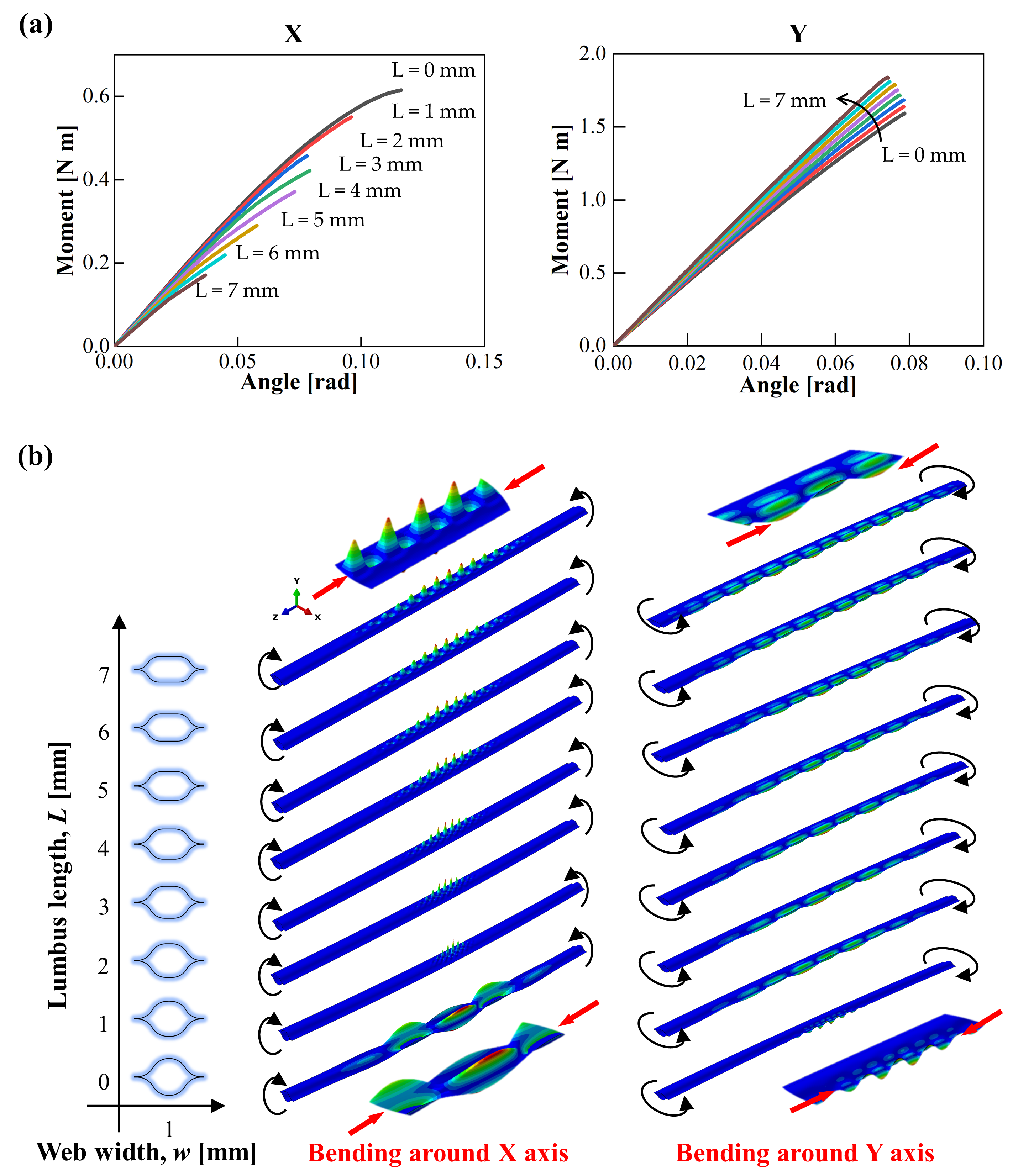}
\caption{Effect of the lumbus length $L$ on (a) the post-buckling response and (b) nonlinear buckling modes of CTLTs with $w=1$ mm when subjected to a pure bending moment about $x-$axis (left) and $y-$axis (right).}
\label{Figure4}
\end{figure}

Next, we take a closer look at the effect of the lumbus length $L$ on the nonlinear buckling behavior of CTLTs by considering a group of samples with $w$ fixed to be 1 mm and $L$ varies from 0 to 7 mm. Fig.~\ref{Figure4}a describes a trade-off between the bending resistance performance about $x-$axis and that about $y-$axis. Specifically, an increase in the lumbus length $L$ leads to a better bending resistance performance about $x-$axis but in the meantime leads to a worse bending resistance performance about $y-$axis. The effect of the lumbus length is also reflected in terms of the nonlinear buckling modes as shown in Fig.~\ref{Figure4}b. Firstly, for the cases the bending is applied about $x-$axis, two distinct nonlinear buckling patterns were observed according to whether the CTLT has lumbus or not. For the CTLT without lumbus, i.e., $L=0$, the buckling mode is a diamond wave pattern, as often found in the buckling of thin-walled cylindrical shells~\cite{fajuyitan2018nonlinear,yadav2019instability,lee2019elastic}. In contrast, for the CTLT with lumbus, i.e., $L>0$, the buckling mode is recognized as "wrinkles" in the lumbus, and the number of the wrinkles increases as the lumbus length $L$ increases from 1 to 7 mm. Secondly, for the cases bending is applied about $y-$axis, the buckling mode is a wave pattern occurs in the compressed web, and the number of waves also increases as the lumbus length increases from 0 to 7 mm. Finally, considering the trade-off between the bending resistance performance about $x-$ and $y-$axes, a value of 3 mm, which is in the middle of the range for the lumbus length $L$, is selected as the optimal one. Note that for the particular cross-sectional shape of CTLT of interest here, the lumbus length has profound effect on the parameter selection and sectional shape design. This effect, however, is omitted in current literature, and it is elucidated for the first time with the aid of constructed design space diagram.

\subsection{Design space for parabolic CTLTs}
\label{BehaviorSpace_ParabolicCTLTs}

\begin{figure}[H]
\centering
\includegraphics[width=150mm]{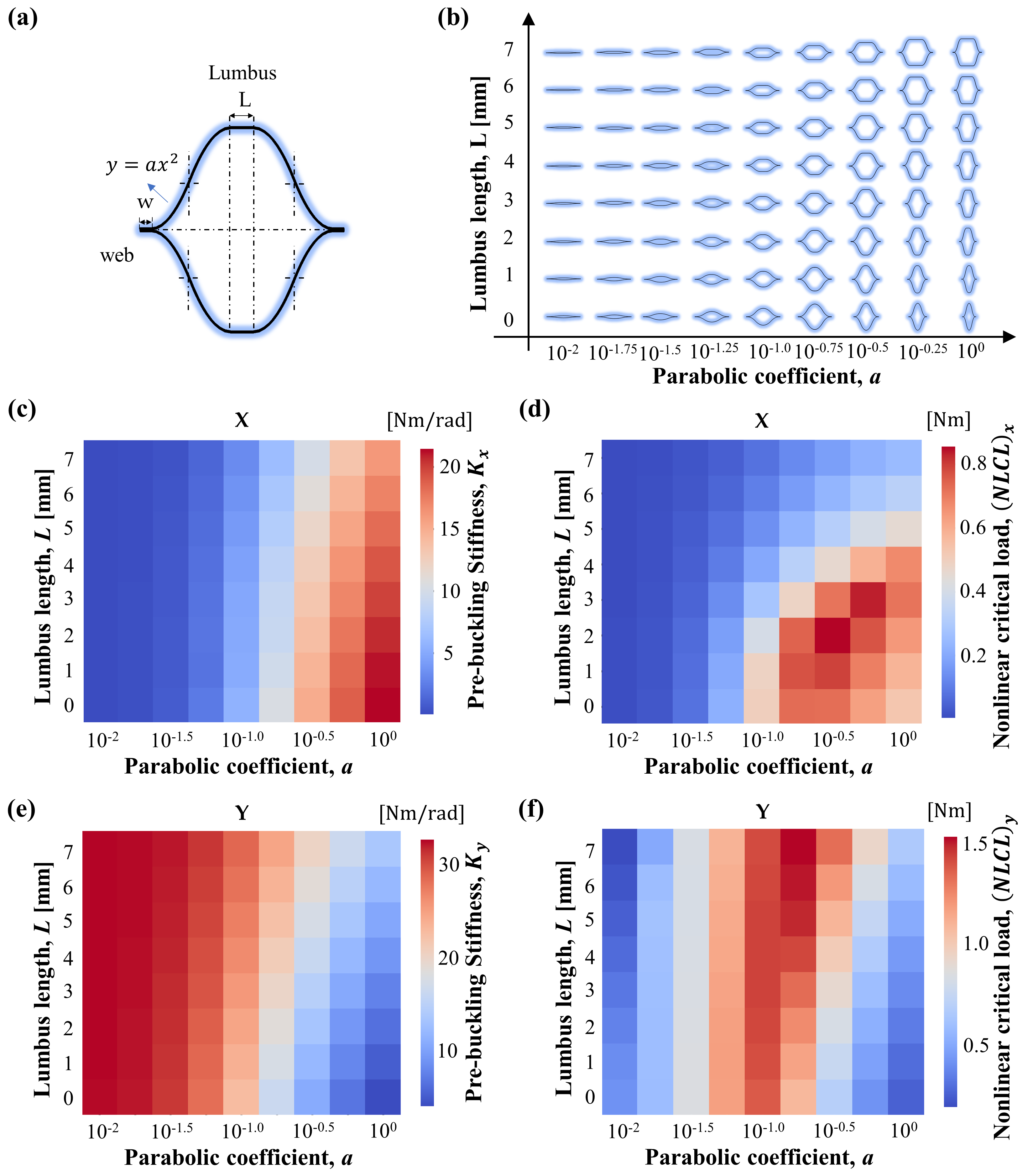}
\caption{The design space for nonlinear buckling of parabolic CTLTs. (a) Schematic illustration of the cross-section geometry of the parabolic CTLT. (b) A variety of cross-section shapes of parabolic CTLTs characterized by the parabolic coefficient $a\in[10^{-2}, 10^0]$ and the lumbus length $L\in[0, 7]$ mm. Heat map illustrating the (c) pre-buckling stiffness $K_x$ and (d) nonlinear critical load $(NLCL)_x$ as a function of the parabolic coefficient $a$ and the lumbus length $L$ when the moment is applied about $x-$axis. Heat map illustrating the (e) pre-buckling stiffness $K_y$ and (f) nonlinear critical load $(NLCL)_y$ as a function of the parabolic coefficient $a$ and the lumbus length $L$ when the moment is applied about $y-$axis.}
\label{Figure5}
\end{figure}

So far, we have explored the nonlinear buckling behavior space of conventional CTLTs with circular arcs, and reported the effect of web and lumbus sizes on the bending performance of circular CTLTs. However, the constant curvature in circular arcs may limit the potential use of conventional CTLTs. A new design of CTLT with variable curvature arcs is achieved by replacing the circular arcs with parabolic arcs, which is referred to as the parabolic CTLT~\cite{lee2018mechanics}. The cross-section geometry of the parabolic CTLT is illustrated in Fig.~\ref{Figure5}a. In this section, we follow our previously developed numerical strategy for parameterizing sectional shape and generating design space for circular CTLTs and extend them to parabolic CTLTs. As shown in Fig.~\ref{Figure5}a, each parabolic arc segment is defined by a parabola function:
\begin{equation}
    y = ax^2 \; \; \; \; 0\leq{x}\leq{x_0}
\end{equation}
where $a$ is the coefficient of the parabola. The arc length of a single parabolic segment, $L_P$, is calculated using the following formula:
\begin{equation}
    L_P = \int_{0}^{x_0}\sqrt{(1+4a^2x^2)}dx
\end{equation}
and then the total flattening length of the cross-section of the parabolic CTLT is expressed as $s=2w+L+4L_P$. Recall that $s$ is set to 27.43 mm as the same as its circular counterpart, and in this case the web width $w$ is fixed to 1 mm. The parabolic CTLT thus has two parameters that can be varied independently, i.e., the lumbus length $L$ and the coefficient of parabola function $a$. Fig.~\ref{Figure5}b presents various cross-section geometries obtained by varying the lumbus length $L$ from 0 to 7 mm and varying the coefficient $a$ from $10^{-2}$ to $10^0$. The coefficient $a$ determines how wide or narrow the parabola is; the greater the coefficient $a$, the narrower the parabola. Therefore, as shown in Fig.~\ref{Figure5}b, the cross-section becomes narrower in $x-$axis and wider in $y-$axis as the coefficient $a$ increases for a given value of lumbus length $L$. In theory, this will cause a decrease in the stiffness about $x-$axis and in the meantime an increase in the stiffness about $y-$axis~\cite{lee2018mechanics,lee2019inducing}. Heat maps in Fig.~\ref{Figure5}c and~\ref{Figure5}d present the effect of the coefficient $a$ and the lumbus length $L$ on the bending performance of parabolic CTLTs about $x-$axis. The pre-buckling stiffness $K_x$ and the nonlinear critical load $(NLCL)_x$ both depend significantly on the coefficient $a$ and the lumbus length $L$. Specifically, a greater coefficient $a$ and/or a smaller lumbus length $L$ leads to an increase in both the pre-buckling stiffness $K_x$ and the nonlinear critical load $(NLCL)_x$. Fig.~\ref{Figure5}e shows the evolution of the pre-buckling stiffness about $y-$axis $K_y$ as a function of the parabolic coefficient $a$ and the lumbus length $L$. In contrast with the performance about $x-$axis, a greater coefficient $a$ and/or a smaller lumbus length $L$ leads to a decrease in the pre-buckling stiffness about $y-$axis $K_y$. In addition, as shown in Fig.~\ref{Figure5}f the parabolic coefficient $a$ has greater effect on the nonlinear critical load $(NLCL)_y$ than the lumbus length $L$, and indicatively, as $a$ increases from $10^{-2}$ to $10^0$, $(NLCL)_y$ increases first, reaches a peak at $a\simeq10^{-1}$, and then decreases.

\begin{figure}[H]
\centering
\includegraphics[width=150mm]{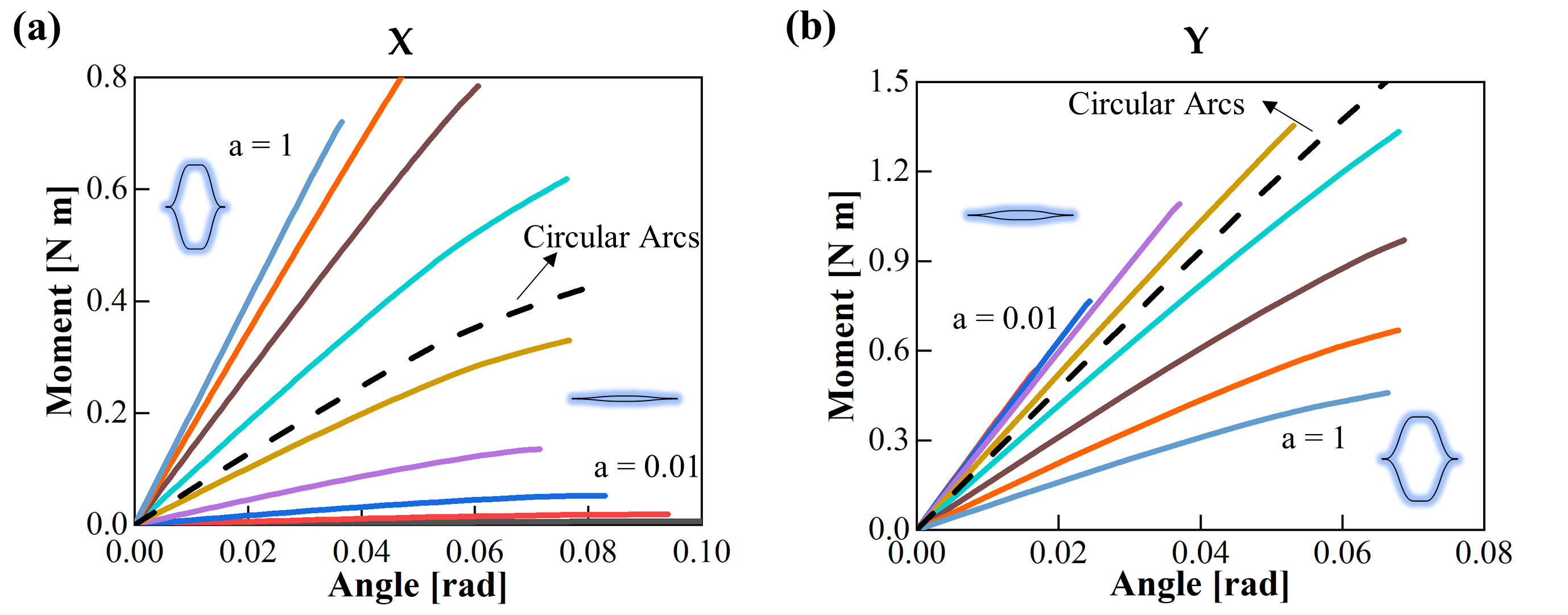}
\caption{Effect of the parabolic coefficient $a$ on the post-buckling response of parabolic CTLTs characterized by $L=3$ mm when subjected to a pure bending moment about (a) $x-$axis and (b) $y-$axis.}
\label{Figure6}
\end{figure}

We also investigated in detail the effect of the parabolic coefficient $a$ on the post-buckling response of parabolic CTLTs by considering a group of samples characterized by $L=3$ mm and $a\in[10^{-2}, 10^0]$. An possible compromise is made once again between the bending resistance performance about $x-$axis and that about $y-$axis. It can be seen from Fig.~\ref{Figure6} that a greater value of parabolic coefficient $a$ leads to a better bending performance about $x-$axis but a worse bending performance about $y-$axis. In addition, CTLTs with circular arcs (black dash lines in Fig.~\ref{Figure6}) show a moderate performance in both cases. These results indicate that a further tunability of the bending performance of CTLTs can be achieved by altering the parabolic coefficient $a$, and it could be useful in real-world engineering applications when anisotropic stiffness property, e.g., a higher stiffness in one direction and a lower stiffness in the other direction, is desirable.

\subsection{Comparison with TRAC booms}
\label{Comparison}
Having explored the design space of circular and parabolic CTLTs, in this subsection, we will perform a comparative study on the bending resistance performance of circular and parabolic CTLTs to that of TRAC booms of equal weight. 

With the design space in hand, the variations of bending performance of both circular and parabolic CTLTs are extracted readily from the heat maps in Fig.~\ref{Figure3} and Fig.~\ref{Figure5}. The same physical quantity is taken from the nonlinear buckling of TRAC booms reported very recently in ~\cite{bessa2018design}. A comparison of variation of the nonlinear critical loads between TRAC booms, circular CTLTs and parabolic CTLTs is given in Table 1. In summary, circular CTLTs has a narrower range of variation $(NLCL)_x$ but a significantly wider range of variation of $(NLCL)_y$ than TRAC booms for the case considered herein. By introducing parabolas, the maximum value of $(NLCL)_x$ of parabolic CTLTs is beyond that of TRAC booms. The maximum value of $(NLCL)_y$ of parabolic CTLTs is slightly lower than that of circular CTLTs, but still much greater than that of TRAC booms.

\begin{table}[H]
\label{table}
\centering
\caption{Comparison of variation of nonlinear critical buckling moment $(NLCL)$ between TRAC booms, circular CTLTs and parabolic CTLTs}
\begin{tabularx}{8cm}{lll}
\hline
                & $(NLCL)_x$ & $(NLCL)_y$\\ \hline
TRAC booms~\cite{bessa2018design}      & $0-0.77$ Nm       & $0-0.31$ Nm       \\
Circular CTLTs  & $0-0.6$ Nm        & $0-2.0$ Nm        \\
Parabolic CTLTs & $0-0.82$ Nm       & $0-1.5$ Nm        \\ \hline
\end{tabularx}
\end{table}

\begin{figure}[H]
\centering
\includegraphics[width=150mm]{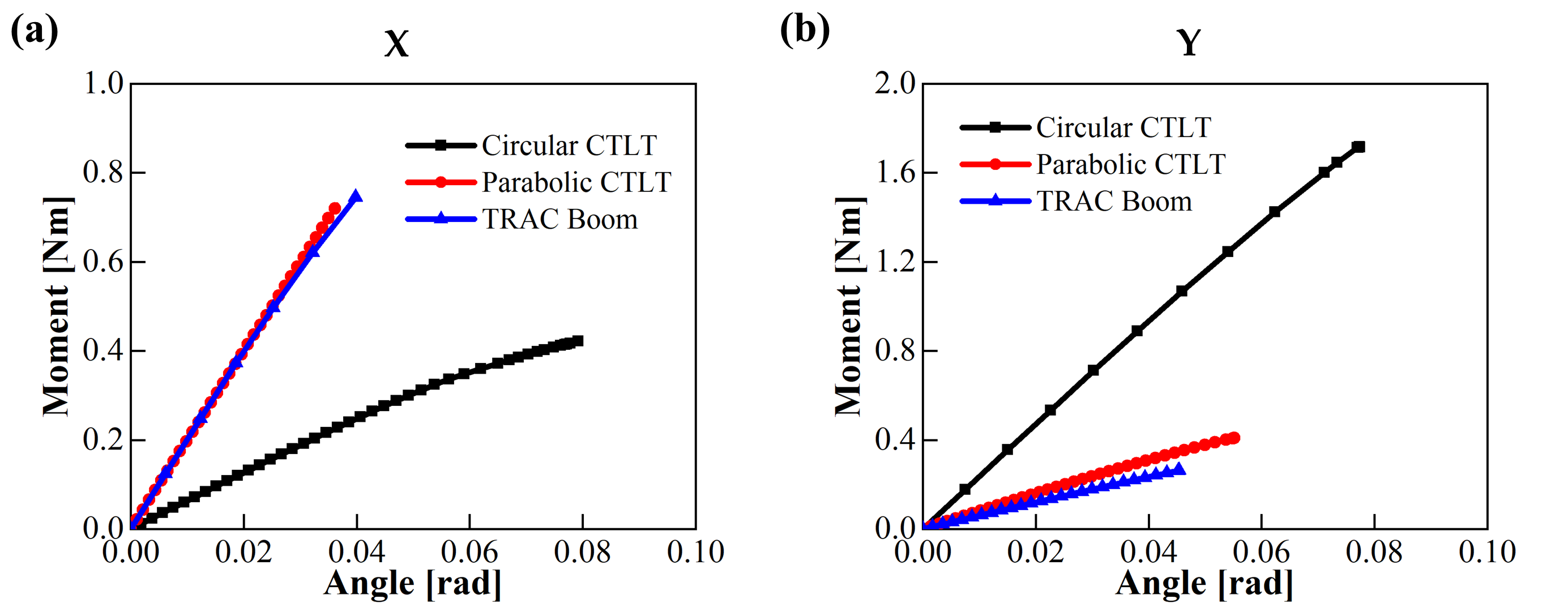}
\caption{Comparison of bending resistance performance around (a) $x-$ axis and (b) $y-$axis between the optimal TRAC boom (blue), circular CTLT (black), and parabolic CTLT (red).}
\label{Figure7}
\end{figure}

In addition, as reported in~\cite{bessa2018design}, in the design space of TRAC booms there is a common region that maximizes the nonlinear critical buckling load for both loading directions at large flange angles and small web height, and this gives rise to an optimal TRAC boom design characterized by a web height of 2 mm and a flange angle of 300$^{\circ}$.  In Fig.~\ref{Figure7} the moment-angle curves of the optimal TRAC boom were reproduced (blue lines) and used as the reference to evaluate the performance of CTLTs. Notice that the results of TRAC boom that reproduced by our approach are in excellent agreement with that reported in~\cite{bessa2018design}, which validate the accuracy of the proposed model. A circular CTLT characterized by $w=1$ mm and $L=3$ mm and a parabolic CTLT characterized by $a=10^0$ and $L=3$ mm are chosen as the optimal candidates from the design space for comparison. As shown in Fig.~\ref{Figure7}, the circular CTLT demonstrates a worse performance than the others in $x-$axis but better in $y-$axis, while the parabolic CTLT shows a comparable performance in both $x$ and $y-$axes to the TRAC boom. These results show that the optimal CTLT with tailored parameters derived from the design space demonstrates a comparable or even better performance than the TRAC boom in either the two loading directions.

\section{Conclusion}
\label{conclusion}

TRAC booms and CTLTs have been proposed as two promising candidates for solutions for deploying various light-weight space structures. Since the thin-walled tubes or booms are thin in thickness and have long aspect ratio, structural instability arises as one of the biggest concerns for structural designs. Recent study on nonlinear buckling analysis of TRAC booms reveals that classical linear buckling may biasedly estimate the critical load of TRAC booms under pure bending and necessitates the classification of nonlinear and linear buckling. However, the recurrence of this difference needs to be justified in CTLTs. Moreover, a design space is desperately needed, with which a thorough exploration and tailoring of design parameters is admissible.

In the present paper, we carried out a comprehensive computational investigation aiming at exploring the design space for nonlinear buckling behavior of CTLTs with both circular and parabolic arcs. Our results reveal the effect of cross-section geometric parameters such as the size of web, lumbus, and the parabolic coefficient, on the bending resistance performance of CTLTs. More specifically, we found that there is a common region at small web width that maximizes the nonlinear critical buckling load of CTLTs concurrently about both $x$ and $y-$axes under pure bending, whereas there is a trade-off between the two loading directions in terms of pre-buckling stiffness, namely, an increase in the web width and/or a descrease in the parabolic coefficient would result in a decrease in the pre-buckling stiffness about $x-$axis but an increase in the pre-buckling stiffness about $y-$axis. More importantly, the contrary effects of the lumbus length on bending performance of CTLTs in two directions are also unveiled. This profound effect of the lumbus, nevertheless, has been ignored in the literature to our best knowledge~\cite{royer2018ultralight}. Moreover, we made a quantitative comparison on the bending performance of TRAC booms and CTLTs under the same weight. The optimal circular CTLT demonstrates a better bending resistance performance than the optimal TRAC boom in $y-$ axis but a worse performance in $x-$axis, while the optimal parabolic CTLT demonstrates a comparable or better performance in both loading directions than TRAC boom.

We believe our efforts provide guidelines for engineers and scientists attempting to design CTLTs with desirable bending resistance performance. We also note that the nonlinear buckling simulation scheme proposed in this work is readily extended for analyzing and optimizing the bending performance of other kinds of thin-walled composite deployable structures~\cite{bessa2018design,yang2019wrapping,yang2020analytical,yang2019coiling}.The paradigm of parameterizing cross-section shape and then constructing a design space in an automatic way can also be developed for optimal design of other thin-walled slender structures. 

\section*{Acknowledgement}

This research is supported by National Natural Science Foundation of China (grant 11972277). Q. J. also acknowledges the support from Shanghai Rising-Star Program (19QB1404000).

\section*{Data availability}
The Abaqus scripts used for the numerical analyses that support the findings of this study are available from the corresponding author upon reasonable request.

\bibliographystyle{elsarticle-num}
\bibliography{main}

\end{document}